\documentclass[letter]{aa} 
\usepackage{graphics,graphicx}
\usepackage[varg]{txfonts}
\usepackage{rotating}
\usepackage{lscape}
\usepackage{rotating}
\usepackage{natbib}
\usepackage{longtable}
\newcommand{\asec}{$^{\prime\prime}$}
\newcommand{\pas}{.\hskip-2pt$^{\prime\prime}$}
\def\IRAS{IRAS16061--5048c1}
\def\I{I16061c1}

\def\H{N$_{2}$H$^{+}$}

\def\CII{\mbox{C$^{18}$O}}

\def\kms{\mbox{km~s$^{-1}$}}
\def\cmc{cm$^{-3}$}
\def\cmq{cm$^{-2}$}

\def\solm{\mbox{M$_\odot$}}

\def\Mvir{$M_{\rm vir}$}
\begin{document}
\title{Magnetically-regulated fragmentation of a massive, dense and turbulent clump}
\author{F. Fontani\inst{1}
            \and
            B. Commer\c{c}on\inst{2} 
            \and
            A. Giannetti\inst{3}
            \and 
            M.T. Beltr\'an\inst{1}
            \and 
            A. S\'anchez-Monge\inst{4}
            \and 
            L. Testi\inst{1,5,6} 
            \and
            J. Brand\inst{7} 
            \and
            P. Caselli\inst{8}
            \and
            R. Cesaroni\inst{1} 
            \and
            R. Dodson\inst{9} 
            \and
            S. Longmore\inst{10} 
            \and
            M. Rioja\inst{9,11,12} 
            \and
            J.C. Tan\inst{13} 
            \and
            C.M. Walmsley\inst{1}
            }
\offprints{F. Fontani, \email{fontani@arcetri.astro.it}}
\institute{INAF-Osservatorio Astrofisico di Arcetri, Largo E. Fermi 5, I-50125, Florence, Italy
           \and 
           Ecole Normale Sup\'erieure de Lyon, CRAL, UMR CNRS 5574, Universit\'e Lyon I, 46 All\'ee d'Italie, 69364, Lyon Cedex 07, France
           \and
           Max-Planck-Institut f\"{u}r Radioastronomie, auf dem H\"{u}gel 69, 53121 Bonn, Germany
           \and
           I. Physikalisches Institut, Universit\"{a}t zu K\"{o}ln, Z\"{u}lpicher Str. 77, 50937 K\"{o}ln, Germany
           \and
           European Southern Observatory, Karl-Schwarzschild-Str 2, D-85748, Garching bei M\"{u}nchen, Germany
           \and 
           Gothenburg Center for Advance Studies in Science and Technology, Chalmers University of Technology and University of Gothenburg, SE-412 96 Gothenburg, Sweden
           \and
           INAF-Istituto di Radioastronomia and Italian ALMA Regional Centre, via P. Gobetti 101, I-40129, Bologna, Italy
           \and
           Max-Planck-Instit\"{u}t f\"{u}r extraterrestrische Physik, Giessenbachstrasse 1, D-85748, Garching bei M\"{u}nchen, Germany
           \and
           International Center for Radio Astronomy Research, M468, University of Western Australia, 35, Stirling Hwy, Crawley, Western Australia, 6009, Australia
           \and
           Astrophysics Research Institute, Liverpool John Moores University, Liverpool, L3 5RF, UK
           \and
           CSIRO Astronomy and Space Science, 26 Dick Perry Avenue, Kensington WA 6151, Australia
           \and
           Observatorio Astronomico Nacional (IGN), Alfonso XII, 3 y 5, E-28014 Madrid, Spain
           \and
           Departments of Astronomy \& Physics, University of Florida, Gainesville, FL 32611, USA
           }
\date{Received date; accepted date}

\titlerunning{Magnetically regulated fragmentation}
\authorrunning{Fontani et al.}

 
\abstract{Massive stars, multiple stellar systems and clusters are born from 
the gravitational collapse of massive dense gaseous clumps, and the way these 
systems form strongly depends on how the parent clump fragments into cores 
during collapse. Numerical simulations show that magnetic fields may be the key 
ingredient in regulating fragmentation. Here we present ALMA observations at
$\sim 0.25$ \asec\ resolution of the thermal dust continuum emission at $\sim 278$~GHz 
towards a turbulent, dense, and massive clump, \IRAS , in a very early evolutionary stage.
The ALMA image shows that the clump has fragmented into many cores along a 
filamentary structure. We find that the number, the total mass and the spatial distribution 
of the fragments are consistent with fragmentation dominated by a strong magnetic field. 
Our observations support the theoretical prediction that the magnetic field plays a 
dominant role in the fragmentation process of massive turbulent clumps.}
  

\maketitle

%

\section{Introduction}
\label{intro}
 
High-mass stars, multiple systems, and clusters are born from the gravitational collapse 
of massive dense clumps (compact structures with $M\geq 100$ \solm , and 
$n$(H$_2$)$\geq 10^4$\cmc ) inside large molecular clouds. Stars more massive than 
8 \solm\ are expected to form either through direct accretion of material in massive cores within 
the clump that does not fragment further (e.g. McKee \& Tan~\citeyear{met}, Tan et al.~\citeyear{tan2013}), 
 or as a result of a dynamical evolution where several low-mass seeds competitively 
 accrete matter in a highly fragmented clump (Bonnell et al.~\citeyear{bonnell2004}).  
 In the latter scenario, each clump forms multiple massive stars and many lower mass stars: 
 the unlucky losers in the competitive accretion competition. There is still vigorous debate 
 on which of these scenarios is more likely to occur, and fragmentation appears to be 
 particularly important in this debate. Theoretical 
 models and simulations show that the number, the mass, and the spatial distribution of the 
 fragments depend strongly on which of the main competitors of gravity is dominant. The 
 main physical mechanisms that oppose gravity during collapse are: intrinsic turbulence, 
 radiation feedback, and magnetic pressure (e.g.~Krumholz~\citeyear{krumholz},
 Hennebelle et al.~\citeyear{hennebelle}). Feedback from nascent protostellar 
 objects through outflows, winds and/or expansion of ionised regions (especially from newly 
 born massive objects) can be important in relatively evolved stages (Bate~\citeyear{bate}), 
 but even then seems to be of only secondary importance (Palau et al.~\citeyear{palau}). 

In a pure gravo-turbulent scenario, the collapsing clump should fragment into many cores, 
 the number of which is comparable to the total mass divided by the Jeans mass 
 (Dobbs et al.~\citeyear{dobbs}); on the other hand, fragmentation can be suppressed 
 by temperature enhancement due to the gravitational energy radiated away from the 
 densest portion of the clump that collapses first (Krumholz~\citeyear{krumholz}), or by 
 magnetic support (Hennebelle et al.~\citeyear{hennebelle}). The work by 
 Commer\c{c}on et al.~(\citeyear{commercon2011}) has shown that models 
 with strong magnetic support predict fragments more massive and less numerous than 
 those predicted by the models with weak magnetic support. The crucial parameter in their 
 3-D simulations is $\mu = (M/\Phi )/(M/\Phi)_{\rm crit}$, where $(M/\Phi )$ is the ratio 
 between total mass and magnetic flux, and the critical value $(M/\Phi)_{\rm crit}$,
  i.e. the ratio for which gravity is balanced by the magnetic field (thus, for $(M/\Phi)_{\rm crit}>1$ 
  the magnetic field cannot prevent gravitational collapse), is given by theory (Mouschovias \& Spitzer~\citeyear{mes}). 
 The outcome of the simulations also depends on other initial global parameters of the 
 clump such as gas temperature, angular momentum, total mass, and average volume 
 density. But once these parameters are fixed, the final population of cores shows a 
 strong variation with $\mu$.
 
 Hiterto, studies of the fragmentation level in massive clumps at the earliest stages 
 of the gravitational collapse remain limited. This investigation is challenging for several 
 reasons: pristine massive clumps are rare and typically located at distances larger than 
 1 kpc, hence to reach the linear resolution required for a consistent comparison with the 
 simulations (about 1000 A.U.) requires observations with sub-arcsecond angular resolution. 
 Furthermore, the small mass of the fragments expected in the simulations (fractions of \solm ) 
 requires extremely high sensitivities. In general, the few studies performed so far with
 sub-arcsecond angular resolution, or close to $\sim 1$\asec , reveal either low fragmentation 
 (e.g.~Palau et al.~\citeyear{palau}, Longmore et al.~\citeyear{longmore}), 
 or many fragments but too massive to be consistent with the gravo-turbulent 
 scenario (Bontemps et al.~\citeyear{bontemps}, Zhang et al.~\citeyear{zhang2015}).
 Furthermore, comparisons with models that assume the actual physical conditions 
 (temperature, turbulence) of the collapsing parent clump have not been published yet.
 
\begin{figure}
\begin{center}
 \includegraphics[angle=0,width=9cm]{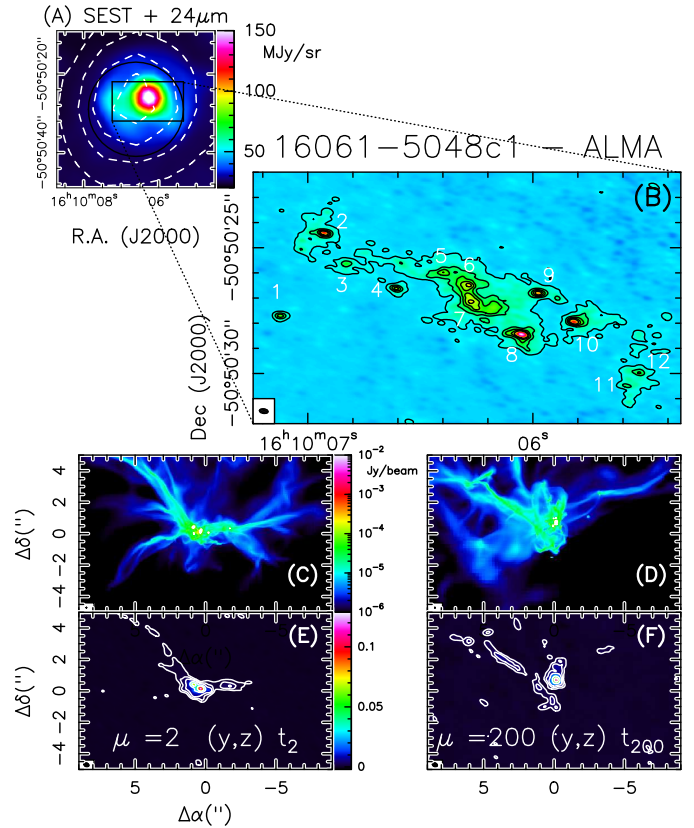}
 \caption[]
 {\label{fig1}{{\bf (A)}: Dust continuum emission map (dashed contours) 
 obtained with the SEST telescope with an angular resolution of 24\asec\ at 
 1.2~mm towards \I\ (Beltr\'an et al.~\citeyear{beltran2006}). The map is 
 superimposed on the Spitzer-MIPS image at 24$\mu$m (in units of MJy/sr). 
 The circle indicates the ALMA primary beam at 278~GHz ($\sim 24$\asec ). 
 {\bf (B)}: Enlargement of the rectangular region indicated in panel {\bf (A)}, showing
the contour map of the thermal dust continuum emission at frequency 278~GHz 
detected with ALMA, in flux density units. The first contour level, and the spacing 
between two adjacent contours, both correspond to the 3$\sigma$ rms of the image 
(0.54~mJy/beam). The cross marks the phase center. The ellipse in the bottom left 
corner shows the synthesized beam, and corresponds to $0.36\times 0.18$\asec\ 
(Position Angle = $86\deg $). The numbers indicate the twelve identified fragments 
(see Sect.~\ref{res}). 
{\bf (C)}: Simulations of the thermal dust emission at 278~GHz 
predicted by the models of Commer\c{c}on et al.~(\citeyear{commercon2011}), 
which reproduce the gravitational collapse of a 300 \solm\ clump, in case of strong 
magnetic support ($\mu=2$), obtained at time t2 (see text), projected on a plane 
perpendicular to the direction of the magnetic field. 
{\bf (D)}: Same as panel {\bf (C)} for the case $\mu=200$ (weak magnetic support). 
{\bf (E)}: Synthetic ALMA images of the models presented in panel {\bf (C)}. 
The contours correspond to 0.54, 1.2, 2, 5, 10, 30, and 50 mJy/beam. 
{\bf (F)}: same as panel (E) for the case $\mu=200$ (weak magnetic support).
}
}
\end{center}
\end{figure}

In this letter, we report on the population of fragments derived in the image 
of the dust thermal continuum emission at $\sim 278$~GHz obtained with the 
 Atacama Large Millimeter Array (ALMA) towards the source \IRAS , hereafter \I , 
a massive ($M\sim 280$\solm , Beltr\'an et al.~\citeyear{beltran2006}, 
Giannetti et al.~\citeyear{giannetti}) and dense (column density of H$_2$, 
$N(H_2)\sim 1.6\times 10^{23}$\cmq ) molecular clump located at 3.6 kpc 
(Fontani et al.~\citeyear{fontani2005}). The clump was detected at 1.2~mm at 
low angular resolution with the Swedish-ESO Submillimeter Telescope 
(SEST, panel (A) if Fig.\ref{fig1}, Beltr\'an et al.~\citeyear{beltran2006}), and
found to be not blended with nearby millimeter clumps, which allows us to avoid 
confusion in the identification of the fragments. Its large mass and column 
density make it a potential site for the formation of massive stars and rich clusters, 
according to observational findings (Kauffmann \& Pillai~\citeyear{kep}, 
Lopez-Sepulcre et al.~\citeyear{lopez}). The clump was classified as an 
infrared dark cloud because undetected in the images of the Midcourse Space 
Experiment (MSX) infrared satellite, although more sensitive images of the 
Spitzer satellite revealed infrared emission at a wavelength of 24$\mu$m (panel (A)
in Fig.~\ref{fig1}). 
Nevertheless, several observational results indicate that the possible embedded 
star formation activity is in a very early stage (Sanchez-Monge et al.~\citeyear{sanchez}). 
In particular, the depletion factor of CO (ratio between expected and observed abundance 
of CO) is 12. This provides strong evidence for the chemical youth of the clump, 
because what causes depletion factors of CO larger than unity is the freeze-out 
of this molecule onto dust grains, a mechanism efficient only in cold and 
dense pre-stellar and young protostellar cores (Caselli et al.~\citeyear{caselli1999}, 
Emprechtinger et al.~\citeyear{emprechtinger}, Fontani et al.~\citeyear{fontani2012}). 
The observations and the data reduction procedures are presented in Sect.~\ref{obs}. 
Our results are shown in Sect.~\ref{res} and discussed in Sect.~\ref{discu}.

\begin{figure}
\begin{center}
\includegraphics[angle=-90,width=10cm]{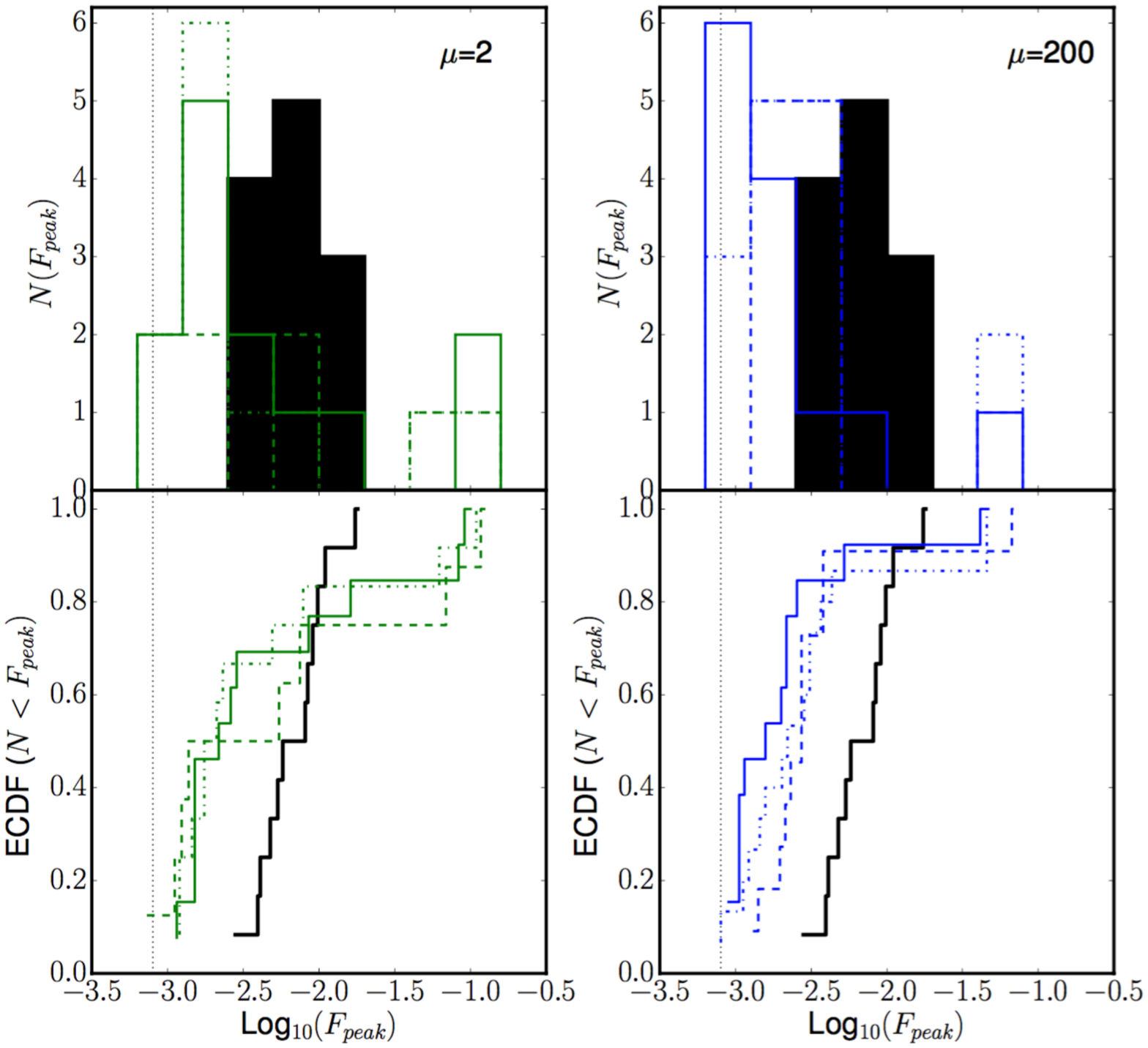}
\caption[]
{\label{fig2}{\it Top panels}: histograms showing the distribution of the peak fluxes 
($F_{peak}$) of the fragments identified in the ALMA image of \I\ (black), and in the 
synthetic images for $\mu$=2 (green, left) and $\mu$=200 (blue, right).  
{\it Bottom panels}: Empirical Cumulative Distribution Function (ECDF) of the quantities 
plotted in the top panels. The black line corresponds to the ALMA data; the green and 
blue lines indicate the strong and weak field cases projected on the 3 planes. 
In all panels, The different line style indicates the projection plane: 
solid = (x,y); dot-dashed = (x,z); dashed = (y,z).
The vertical dotted line corresponds to 0.8 mJy, which is approximately 
5 times the rms noise level in both the real and synthetic maps. 
Note that the $\mu$=2 model spans the observations, while the $\mu$=200 model 
is strongly biased towards fragments with masses lower than those observed.}
\end{center}
\end{figure} 

\section{Observations and data reduction}
\label{obs}

Observations of \I\ with the ALMA array were performed during
southern winter, 2015. The array was in configuration C36-6, with maximum 
baseline of 1091 m. The phase centre was at R.A. (J2000): 16$^h$10$^m$06\pas 61 
and Dec (J2000): $-50^{\circ}50^{\prime}29$\arcsec . The total integration time on 
source was $\sim 18$ minutes. The precipitable water vapour during observations 
was $\sim 1.8$ mm. Bandpass and phases were calibrated by observing
J1427--4206 and J1617--5848, respectively. The absolute flux scale was
set through observations of Titan and Ceres.
From Beltr\'an et al.~(\citeyear{beltran2006}), we know that the 
total flux measured with the single-dish in an area corresponding to the ALMA primary 
beam at the observing frequency (24\asec ) is $\sim 2.3$ Jy, while the total flux 
measured with ALMA in the same area is 0.63 Jy. Thus, we recover $\sim 30\%$ 
of the total flux. The remaining $\sim 70\%$ is likely contained in an extended 
envelope that is resolved out. Continuum was extracted by averaging in frequency 
the line-free channels. The total bandwidth used is $\sim 1703$~MHz.
Calibration and imaging was performed with the CASA\footnote{The Common Astronomy 
Software Applications (CASA) software can be downloaded at http://casa.nrao.edu} 
software (McMullin et al. 2007), and the final images were analysed following 
standard procedures with the software MAPPING of the GILDAS\footnote{http://www.iram.fr/IRAMFR/GILDAS} 
package.
The angular resolution of the final image is $\sim 0.25$ \asec\ (i.e.$\sim 900$ AU at the source distance). 
We were sensitive to point-like fragments of 0.06 \solm .
Together with the continuum, we detected several lines among which \H (3--2).
These data will be presented and discussed in a forthcoming paper. In this
letter, we only use the \H\ (3--2) line to derive the virial masses, as we will show
in Sect.~\ref{res}.

\section{Results}
\label{res}
B
The ALMA map of the dust thermal continuum emission is shown in 
Fig.~\ref{fig1}{\bf (B)}: we have detected several dense condensations distributed 
in a filamentary-like structure extended east-west, surrounded by fainter 
extended emission. This structure has been decomposed into twelve fragments 
(Fig.~\ref{fig1}{\bf (B)}). The fragments have been identified following these criteria: 
(1) peak intensity greater than 5 times the noise level; (2) two partially overlapping 
fragments are considered separately if they are separate at their half peak intensity 
level. The minimum threshold of 5 times the noise was adopted according to the fact 
that some peaks at the edge of the primary beam are comparable to about 4-5 
times the noise level. We decided to use these criteria and decompose the map 
into cores by eye instead of using decomposition algorithms (such as Clumpfind) 
because small changes in their input parameters could lead to big changes in 
the number of identified clumps (Pineda et al.~\citeyear{pineda}).
The main physical properties of the fragments derived from the continuum map, 
i.e. integrated and peak flux, size, and gas mass, and the methods used to
derive them, are described in Appendix~A. The derived parameters 
are listed in Table~\ref{tab1}.
The mean mass of the fragments turns out to be 4.4 \solm , with a minimum value 
of 0.7 \solm\ and a maximum of $\sim 9$ \solm . The diameters (undeconvolved
for the beam) range from 0.011 to 0.032~pc, with a mean value of 0.025~pc. 

To investigate the stability of the fragments, we have calculated the virial masses \Mvir , 
i.e. the masses required for the cores to be in virial equilibrium, from the line widths 
observed in \H\ (3--2). As stated in Sect.~\ref{obs}, in this work we use this molecular 
transition only for the purpose to derive the level of turbulence (the key ingredient of 
the models, together with the magnetic support) of the dense gas out of which the 
fragments are formed. The approach used to derive \Mvir\ is described in Appendix~A
and the values obtained are reported in Table~\ref{tab1}. 
The average ratio between \Mvir\ and $M$ computed from the continuum emission is 
about 0.4, indicating that the gravity dominates, according to other ALMA studies of
fragmentation (Zhang et al.~\citeyear{zhang2015}). However, the uncertainties due to 
the dust mass opacity coefficient (see Eq.~\ref{mass_dust}) can be of a factor 2-3, 
hence it is difficult to conclude that the fragments are unstable. Moreover, the formula of
the virial mass we use does not consider the magnetic support. Because this latter is expected 
to be relevant, it is likely that the fragments are closer to virial equilibrium and would not 
tend to fragment further. If one assumes, for example, the value of 0.27~mG measured by 
Pillai et al.~(\citeyear{pillai2015}) in another infrared-dark cloud, the ratio between virial 
mass and gas mass becomes about 1, and the fragments would be marginally stable. 
A similar conclusion is given in Tan et al.~(\citeyear{tan2013}), were the dynamics of 
four infrared-dark clouds similar to \I\ is performed. 

\section{Discussion and conclusion}
\label{discu}

We have simulated the gravitational collapse of \I\ through 3D 
numerical simulations following Commer\c{c}on et al.~(\citeyear{commercon2011}), 
adopting mass, temperature, average density, 
and level of turbulence of the parent clump very similar to those measured (Beltr\'an
et al.~\citeyear{beltran2006}, Giannetti et al.~\citeyear{giannetti}). In particular, the Mach 
number setting the initial turbulence, has been derived from the line width of 
\CII\ (3--2) by Fontani et al.~(\citeyear{fontani2012}). Because these observations were 
obtained with angular resolution of $\sim 24$\asec , 
and the critical density of the line ($\sim 5\times 10^4$\cmc , Fontani et al.~\citeyear{fontani2005}) 
is comparable to the average density of the clump as a whole (Beltr\'an et al.~\citeyear{beltran2006}), 
the \CII\ line width represents a reasonable estimate of the intrinsic turbulence of the 
parent clump. The models were run for $\mu$=2 (strongly magnetised case) 
and $\mu$=200 (weakly magnetised case). Then, we have post-processed the 
simulations data and computed the dust emission maps at 278~GHz: the final 
maps obtained in flux density units at the distance of the source have been 
imaged with the CASA software, in order to reproduce synthetic images 
with the same observational parameters as those of the observations. 
A detailed description of the parameters used for the numerical simulations, 
of the resulting maps, and how they have been post-processed, is given in 
Appendix~B. Further descriptions of the models can be found also in Commer\c{c}on
et al.~(\citeyear{commercon2011}).
To investigate possible effects of geometry, we have imaged the outcome of 
the simulations projected on three planes: (x,y), (x,z) and (y,z), where x is the 
direction of the initial magnetic field. As an example, in Figure~\ref{fig1}{\bf (C)} 
and \ref{fig1}{\bf (D)} we show the results for the cases of strong and weak magnetic 
support, respectively, projected on the (y,z) plane, i.e. on a direction perpendicular to the 
magnetic field. The synthetic images obtained with CASA are shown in panels
{\bf (E)} and {\bf (F)}. All the planes for both $\mu$=2 and $\mu$=200 are shown in 
Appendix~B, in Figs.~\ref{fig_appb1} and \ref{fig_appb2}, respectively.
An important result of the simulations (see Figure \ref{fig_appb3} in Appendix B) is that the 
total flux seen by the interferometer in the $\mu$=2 case decreases until about 
35$\times 10^3$ yrs and then goes through a minimum and starts gradually to increase. 
In the $\mu$=200 case by contrast, the decrease is not reversed. We conclude from this 
that in the $\mu$=2 case, the fragments continue to accrete material and eventually 
they will reach the total flux observed in the ALMA image. Thus, we have in the 
$\mu$=2 case analysed the synthetic images produced at the time at which the total 
flux in the fragments matches the observed value within an uncertainty of about 
10$\%$ (the calibration uncertainty on the flux density, see the ALMA Technical 
Handbook\footnote{https://almascience.eso.org/proposing/technical-handbook}, 
while for the $\mu$=200 case we have analysed the synthetic images obtained at 
the end of the simulations. This corresponds to two different times: t$_2$ = 48500 yrs after 
the birth of the first protostar for $\mu$=2; t$_{200}$ = 59500 yrs after the birth of the first 
protostar for $\mu$=200. 

The synthetic maps with $\mu$=200 show more fragments with lower peak fluxes, 
and the overall structure is more chaotic and never filamentary, independently 
of the projection plane. The identification of the fragments and the derivation of 
their properties in the synthetic images have been made following the same criteria 
and procedures used for the ALMA data. Therefore, any systematic error introduced 
by the assumptions made (e.g. the assumed dust temperature, gas-to-dust ratio, 
dust grain emissivity) are the same both in the real and synthetic images, thus they 
do not affect their comparison. The statistical properties of the synthetic core populations 
are summarised in Table~\ref{tab2} of Appendix~B. We have also compared the 
cumulative distribution of the peak fluxes of the fragments in the observed and 
synthetic images. The results are shown in Fig.~\ref{fig2}. The case with $\mu$=200 
has lower peak values for the whole populations, while 1 or 2 fragments have a 
higher peak value than the maximum observed. The $\mu$=2 case has a broader 
distribution of values.
Overall, the ALMA map shows a narrower distribution of peak fluxes, with a deficit 
of both very weak and very strong peaks, which in turn are present in both synthetic 
images. However, the $\mu$=2 model roughly spans the observations, while the 
$\mu$=200 model is heavily biased below the data. Also, non-parametric statistical test 
(Anderson-Darling test) implies that all the $\mu$=200 cases can be excluded as being 
drawn from the same parent distribution as the observed values with a confidence level 
exceeding 99.8$\%$. The $\mu$=2 case is less obvious, because two projections could 
be excluded at a 98-99$\%$ confidence level, while the third projection, (y,z), with a 
null hypothesis probability of $\sim 90\%$ cannot be excluded at the 2$\sigma$ level. 
The deficit of very strong and very weak peaks in the real image may be due to a 
difference between the $\mu$ values assumed in the simulations and the real one, 
or to some other unknown (or doubtful) initial assumption such as, e.g., the density 
profile or the homogeneous temperature of the collapsing clump.

\vspace{0.3cm}
\noindent
Based on the overall morphologies shown in Fig.~\ref{fig1} (and in Figs.~\ref{fig_appb1} and \ref{fig_appb2}), 
and on the statistical properties of the fragments reported in Fig.~\ref{fig2}, undoubtedly 
the model that better reproduces the data is the one with $\mu$=2. Hence, with these new ALMA 
observations, compared with realistic 3D simulations that assume as initial conditions the 
properties of the parent clump, we demonstrate that the fragmentation due to self-gravity 
is dominated by the magnetic support, based on the evidence that: (1) the overall 
morphology of the fragmenting region is filamentary, and this is predicted only in case 
of a dominant magnetic support; (2) the observed fragment mass distribution is most 
easily understood in simulations assuming substantial magnetic support. 

\vspace{0.4cm}
\noindent
{\it Acknowledgments.} 
This paper makes use of the following ALMA data: ADS/JAO.ALMA.2012.1.00366.S. 
ALMA is a partnership of ESO (representing its member states), NSF (USA) and NINS 
(Japan), together with NRC (Canada), NSC and ASIAA (Taiwan), and KASI (Republic of Korea), 
in cooperation with the Republic of Chile. The Joint ALMA Observatory is operated by ESO, 
AUI/NRAO and NAOJ. We acknowledge the Italian-ARC node for their help in the reduction 
of the data. We acknowledge partial support from Italian Ministero dell'Istruzione, Universit\'a 
e Ricerca through the grant Progetti Premiali 2012 $-$ iALMA (CUP C52I13000140001) 
and from Gothenburg Centre of Advanced Studies in Science and Technology through 
the program {\it Origins of habitable planets}.

{}

\clearpage

\renewcommand{\thetable}{A-\arabic{table}}
\renewcommand{\thesubsection}{A-\arabic{subsection}}
\setcounter{table}{0}
\section*{Appendix A: Physical properties of the fragments}
\label{appa}

\subsection{Derivation of the parameters}
\label{appa_parameters}

\begin{itemize}

\item{{\bf Integrated flux densities}:
The integrated flux densities of the fragments, $S_{\nu}$, have been obtained from the 3$\sigma$ rms 
contour in the continuum image. In the few cases for which the 3$\sigma$ rms contours of two
adjacent fragments are partly overlapping (e.g.~fragments 5, 6, and 7 in Fig.~\ref{fig1}), 
the edges between the two have been defined by eye at approximately half of the separation 
between the peaks. The results are shown in 
Table~\ref{tab1}.}

\item{{\bf gas masses}: The gas mass of each fragment has been calculated from the equation:
\begin{equation}
M=\frac{g S_{\nu} d^2}{\kappa_{\nu}B_{\nu}(T_{\rm d})}\;,
\label{mass_dust}
\end{equation}
        
where $S_{\nu}$ is the integrated flux density, $d$ is the distance to the source, $\kappa_{\nu}$ 
is the dust mass opacity coefficient, $g$ is the gas-to-dust ratio (assumed to be 100), 
and $B_{\nu}(T_{\rm d})$ is the Planck function for a black body of temperature $T_{\rm d}$. 
We adopted $T_{\rm d}=25$ K, corresponding to the gas temperature derived by Giannetti
et al.~(\citeyear{giannetti}), assuming coupling between gas and dust 
(reasonable assumption at the high average density of the clump). The dust mass opacity 
coefficient was derived from the equation $\kappa_{\nu}=\kappa_{\nu_0}(\nu/\nu_0)^{\beta}$. 
We assumed $\beta=2$ and $\kappa_{\nu_0}=0.899$ \cmq\ gr$^{-1}$ at $\nu_0 = 230$~GHz, 
according to Ossenkopf \& Henning~(\citeyear{oeh}). The largest mass derived is ~9 \solm , 
the smallest is $\sim 0.7$ \solm\ (see Table~\ref{tab1}).}

\item{{\bf Size}: the size of each fragment has been estimated as the diameter of the circle with 
area equivalent to that encompassed by the 3$\sigma$ rms contour level. The results are shown in Table~\ref{tab1}.
The ALMA beam size is much smaller than the size of the fragments, so that deconvolution
for the beam is irrelevant to derive the source size.}

\item{{\bf Virial masses}: the virial masses were derived in this way: first, we extracted the \H\ (3--2) spectra 
from the 3$\sigma$ rms level of the 12 continuum cores. All the spectra are fitted in 
an automatic way using a procedure based on the integration of the python module 
PyMC  and the CLASS extension WEEDS (Maret et al.~\citeyear{maret}). 
Then, the virial masses, \Mvir , are computed from the formula
\begin{equation} 
M_{\rm vir} = 210 r \Delta v^{2} \solm \;,
\end{equation}
where $r$ is the size of the fragment, and $\Delta v$ is the line width at half maximum of the 
average \H\ (3--2) spectrum obtained from the fitting procedure described above. 
The results are shown in Table~\ref{tab1}.}

\end{itemize}

\addtocounter{table}{0}

\begin{table*}
\begin{center}
\caption[] {Peak position (in R.A. and Dec. J2000), integrated flux (inside the 3$\sigma$ rms contour level), 
peak flux, diameter, mass, line width at half maximum, and virial mass of the 
12 fragments identified in Fig.~\ref{fig1}{\bf (B)}. The line widths are computed from 
the \H\ (3--2) spectra extracted from the polygons defining the external profile of each 
fragment, as explained in Appendix A.}
\label{tab1}
\begin{tabular}{lcccccccc}
\hline \hline
Fragment	& R.A. (J2000) & Dec. (J2000) & $S_{\nu}$ & $S_{\nu}^{\rm peak}$ & $D$ & $M$  & $\Delta v$ & \Mvir\ \\
        & h:m:s      &    $\deg:\prime$:\asec & Jy     & Jy beam$^{-1}$ & pc & \solm\ & \kms\ & \solm\ \\
\hline
1	& 16:10:07.12	& $-$50:50:27.7 &	0.007 &	0.0054 &	0.011 &	0.72 &	0.51 &	0.28 \\
2	& 16:10:06.93	& $-$50:50:24.4 &	0.045 &	0.0092 &	0.031 &	4.70 &	0.90 &	1.86 \\
3	& 16:10:06.83 & $-$50:50:25.6 &	0.012 &	0.0041 &	0.018 &	1.25 &	0.48	&      0.47 \\
4	& 16:10:06.60 & $-$50:50:26.6 &	0.012 &	0.0058 &	0.016 &	1.25	&      0.82 &	1.26 \\
5	& 16:10:06.40	& $-$50:50:26.0 &	0.044 & 	0.0048 &	0.029 &	4.59	&      0.49	&      0.77 \\
6	& 16:10:06.29	& $-$50:50:26.5 &	0.073 &	0.0086 &	0.028 &	7.62 &	0.36 &	0.38 \\
7	& 16:10:06.27	& $-$50:50:27.1 &	0.050 &	0.0083 &	0.023 &	5.22	&      0.33	&      0.26 \\
8	& 16:10:06.16	& $-$50:50:27.3 &	0.052 &	0.0180 &	0.024 &	5.42 &	0.84 &	2.01 \\
9	& 16:10:05.98	& $-$50:50:26.8 & 	0.045 &	0.010  &	0.029 &	4.70 &	0.33	&      0.33 \\
10	& 16:10:06.05	& $-$50:50:28.4 &	0.084 &	0.011  &	0.031 &	8.76 &	1.04 &	3.47 \\
11	& 16:10:05.82	& $-$50:50:27.9 &	0.052 &	0.0028 &	0.032 &	5.43 &	1.00 &	3.29 \\
12	& 16:10:05.53	& $-$50:50:30.0 &	0.032 &	0.0041 &	0.030 &	3.34 &	0.76 &	1.82 \\
\hline
\end{tabular}
\end{center}
\end{table*}



\renewcommand{\thefigure}{B-\arabic{figure}}
\renewcommand{\thetable}{B-\arabic{table}}
\renewcommand{\thesubsection}{B-\arabic{subsection}}
\setcounter{table}{0}
\setcounter{figure}{0}
\setcounter{subsection}{0}
\section*{Appendix B: Simulations and synthetic images}
\label{appb}

\subsection{Methods and initial conditions for the numerical calculations}
\label{appb1}

We perform a set of two radiation-magneto-hydrodynamics calculations 
which includes the radiative feedback from the accreting protostars. 
We use the RAMSES code with the grey flux-limited-diffusion approximation 
for radiative transfer and the ideal MHD for magnetic fields (Commer\c{c}on
et al.~\citeyear{commercon2012},~\citeyear{commercon2014}, Teyssier~\citeyear{teyssier},
Fromang et al.~\citeyear{fromang}). The initial conditions are similar to those 
used in Hennebelle et al.~(\citeyear{hennebelle}) and Commer\c{c}on et 
al.~(\citeyear{commercon2011}) with slight modifications in order to match roughly 
the observed properties of \I . Note that the models presented in this paper have 
been made with initial conditions very similar to those measured from previous 
observations in \I , to perform an appropriate comparison with observations for this 
specific source. Our aim is not to fine-tune the initial conditions such that the models 
best reproduce the observations.  
We consider an isolated spherical core of mass $300$ \solm , radius ~0.25 pc and temperature 
20 K. We assume a Plummer-like initial density profile $\rho (r)=\rho_c/(1+(r/r_0)^2)$, with 
$\rho_c=3.96\times 10^5$ \cmc\ and $r_0=0.085$ pc, and a factor 10 for the density 
contrast between the center and the border of the core. Such a density profile is suggested
by observational findings (Beuther et al.~\citeyear{beuther}). 
The initial magnetic field is aligned with the x-axis and its intensity is proportional to the 
column density through the cloud (Hennebelle et al.~\citeyear{hennebelle}). 
In this paper, we investigate two degrees of magnetization, $\mu$=2 which is close 
to the values 2-3 that are observationally inferred (e.g., Crutcher~\citeyear{crutcher}), 
and $\mu$=200, which corresponds to a quasi-hydrodynamical case. 
Last, we apply an initial internal turbulent velocity field to the cores. The velocity field 
is obtained by imposing a Kolmogorov power spectrum with randomly determined 
phases (i.e., a ratio 2:1 between the solenoidal and the compressive modes). 
There is no global rotation of the cloud, meaning that the angular momentum 
is contained within the initial turbulent motions, which are then amplified by the 
gravitational collapse. The amplitude of the velocity dispersion is scaled to match 
a turbulent Mach number of 6.44, in agreement with \CII\ observations (Fontani et 
al.~\citeyear{fontani2012}). Following Hennebelle et al.~(\citeyear{hennebelle}, 
Eq. 2 therein), the virial parameter is $\alpha_{\rm vir}\sim 0.72$ for $\mu$=2, 
and $\alpha_{\rm vir}\sim 0.54$ for $\mu$=200 (close to virial equilibrium in both cases). 
The two calculations, $\mu$=2 and $\mu$=200, start with the same initial turbulent 
velocity field (only one realisation is explored) and turbulence is not maintained 
during the collapse. Investigation of the effect of different initial turbulent seeds 
is beyond the scope of this paper. The computational box has a 2563 resolution, 
and the grid is refined according to the local Jeans length (at least 8 cells/Jeans length) 
down to 7 levels of refinement (minimum cell size of 13 AU). Below 13 AU, 
the collapsing gas is described using sub-grids models attached to sink particles, 
similar to what is done in other studies (Krumholz et al.~\citeyear{krumholz2009}). 
We use the sink particle method presented in (Bleuler et al.~\citeyear{bleuler}), 
though with slight modifications on the checks performed for the sink creation. 
The sink particles accrete the gas that sit in their accretion volumes (sphere of radius $\sim 52$ AU, 
4 cells) and that is Jeans unstable. We consider that half of the mass accreted into the 
sink particles actually goes into stellar material. The luminosity of the protostars is then 
computed using mass-radius and mass-luminosity empirical relations of main sequence stars 
(e.g.~Weiss et al.~\citeyear{weiss}) and is injected within the accretion volume in the 
computational domain as a source term (e.g.~Krumholz et al.~\citeyear{krumholz2009}). 
We do not account for accretion luminosity. 

\subsection{Outcomes of the numerical calculations}
\label{appb2}

We run the calculations until they reach a star formation efficiency (SFE) > $20\%$ 
(where the star formation efficiency corresponds to the ratio between the mass of the 
gas accreted into the sink particles and the total mass of the cloud). Again, the choice 
of the times at which we stop the calculations is not aimed at best reproducing the observed values. 
Model $\mu$=2 is post-processed at time t$_2$ = 48500 yrs after the birth of the 
first protostar, which is the time at which the total flux in fragments is equal to the 
observed value (within the uncertainty), and  $\mu$=200 at time t$_200$ = 59500 yrs. 
At these times, model $\mu$=2 has formed 38 sink particles (for a total mass of 60 \solm ) 
while model $\mu$=200 has formed 119 sink particles (for a total mass of 85 \solm ). 
Fig.~\ref{fig_appb3} shows the time evolution after the first sink creation of: the SFE, 
the number of sinks, and the total flux at 278 GHz (within a total area of 80000 AU $\times 80000$ AU) 
for the two models. The circles indicate the time at which the simulations are post-processed. 
Note the increase in the 278 GHz flux with time for $\mu$=2, which probably reflects the 
temperature increase caused by the radiation of the sink particles.

\subsection{Production of the synthetic images}
\label{appb3}

\begin{figure}
\begin{center}
\includegraphics[angle=90,width=9cm]{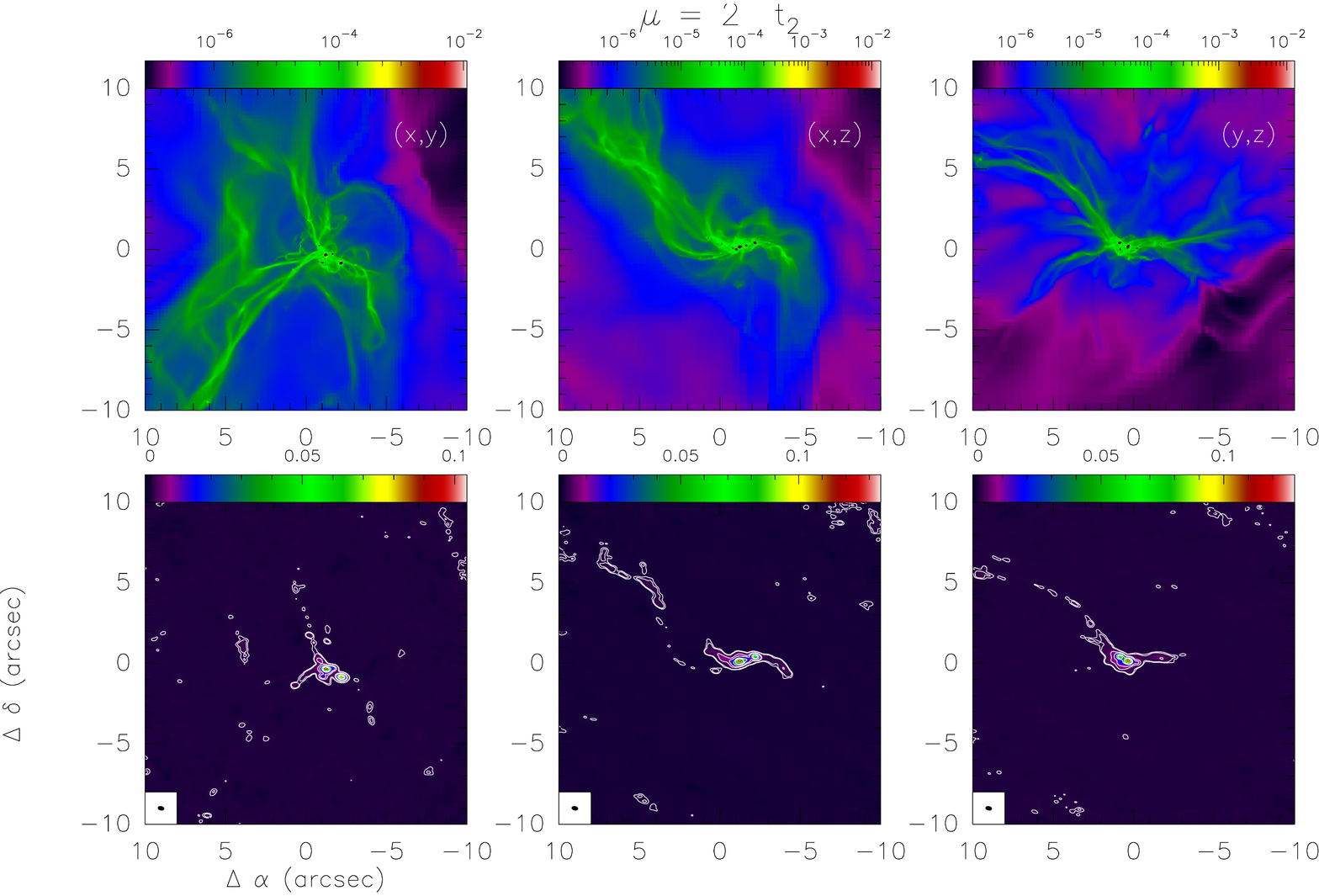}
\caption[]
{\label{fig_appb1} Top panels show the thermal dust continuum emission map 
at frequency 278 GHz predicted by the models of Commer\c{c}on et al.~(\citeyear{commercon2011}), 
which reproduce the gravitational collapse of a 300 \solm\ clump, in case of strong magnetic 
support ($\mu$=2) at time t$_2$ = 34300 years after the birth of the first protostar 
(see main text for details). In the bottom panels, we show the models after processing 
in the CASA simulator, adopting the same observational conditions of the real observations. 
Units of the colour-scale are Jansky/beam. Contour levels are 0.6, 1, 5, 10, 30 and 50 
mJy beam$^{-1}$ in all bottom panels.
}
\end{center}
\end{figure}

\begin{figure}
\begin{center}
\includegraphics[angle=90,width=9cm]{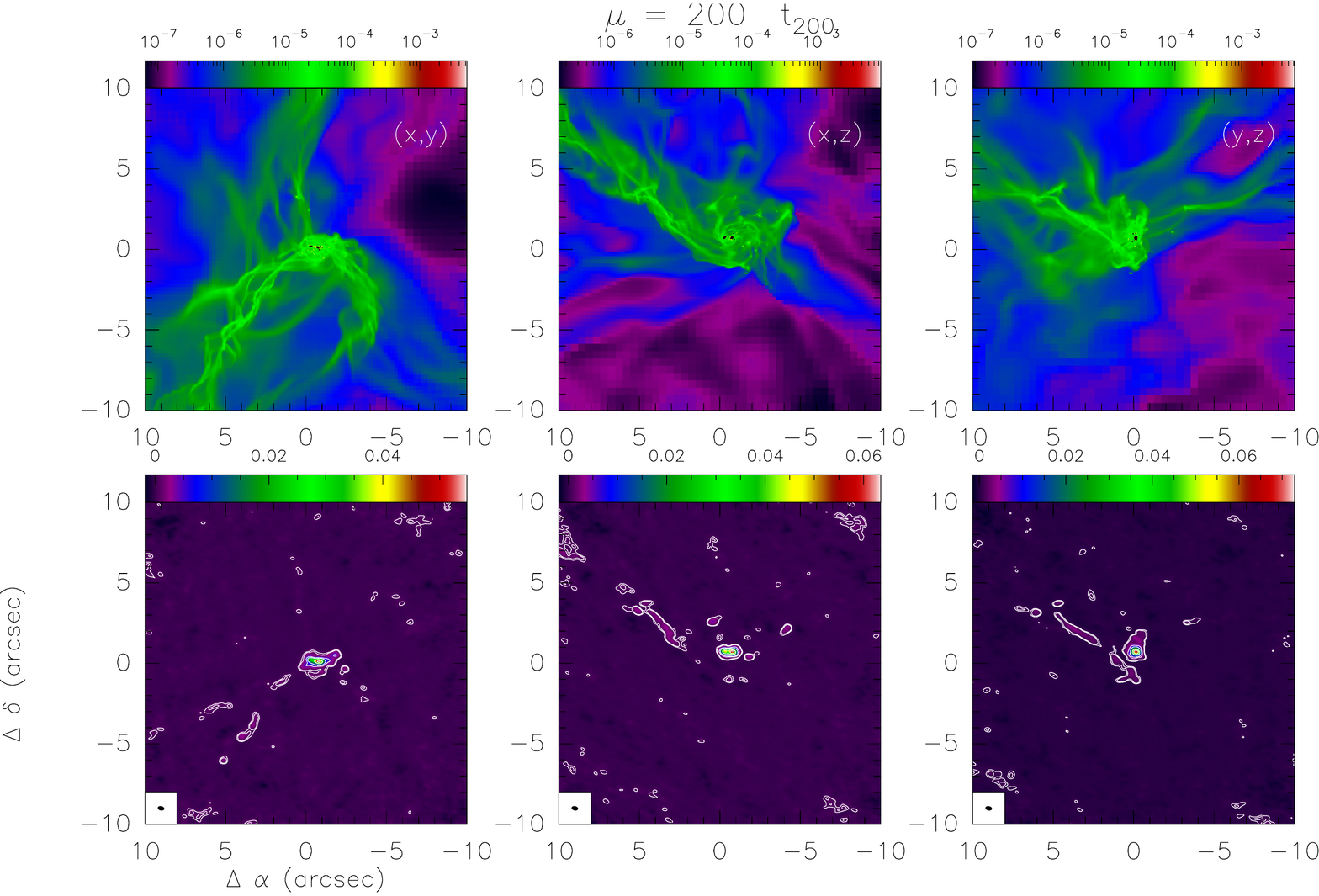}
\caption[]
{\label{fig_appb2}
Same as Fig.~\ref{fig_appb1} for the case $\mu$=200 at time t$_{200}$ = 59500 yrs after 
the birth of the first protostar (see main text for details).
}
\end{center}
\end{figure}

\begin{figure}
\begin{center}
\includegraphics[angle=0,width=9cm]{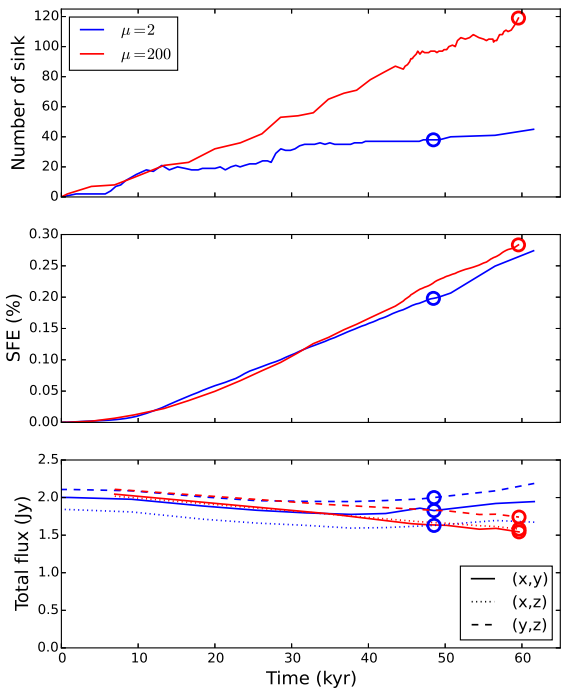}
\caption[]
{\label{fig_appb3}{\it From top to bottom:} evolution with time of the number of sink particles, 
of the SFE, and of the total flux emission at 278 GHz (within a total area of 80000 AU $\times$ 80000 AU) 
for the two models after the creation of the first sink. The circles indicate the time at which the 
simulations are post-processed. In the bottom panel, the different lines correspond to the 
different projection planes as illustrated in the bottom-right corner.
}
\end{center}
\end{figure}

We first postprocessed the RAMSES calculations results using the radiative transfer 
code RADMC-3D\footnote{http://www.ita.uni-heidelberg.de/~dullemond/software/radmc-3d/} 
and the interface presented in (Commer\c{c}on et al.~\citeyear{commercon2012}). 
We produced dust emission maps at 278 GHz (see Figs.~\ref{fig_appb1} and \ref{fig_appb2}). 
We do not account for the stellar luminosities in the synthetic images since the 
stellar irradiation is reprocessed in the envelope at millimetre wavelengths. 
However, we attempted to create models accounting for protostellar luminosities, 
and found results that do not change significantly at the wavelength considered. 
The ALMA synthetic images of the numerical simulations have been then produced 
through the CASA software: first, synthetic visibilities have been created with the task 
{\it simobserve}, which have then been imaged with the task {\it simanalyze}. 
To precisely reproduce the observations, in the tasks we have used the same 
parameters of the observations: integration time on source of 18 minutes, 
precipitable water vapour of about 1.8 mm, array configuration C36-6, start 
hour angle of 2.4 hours (see Sect.~\ref{obs}). 
The population of fragments in the final synthetic images were derived 
following the same procedure described in Sect.~\ref{discu}.

\begin{table}
\begin{center}
\caption[] {Statistical comparison between the fragment population derived from the ALMA image 
of \I\ shown in Fig.~\ref{fig1} and the simulations presented in Figs.~B1 and B2 of the Appendix. 
The derivation of the parameters obtained for both the observed and synthetic images is described 
in Sect.~\ref{res} and in Appendix A, respectively.}
\label{tab2}
\small
\begin{tabular}{lcccccc}
\hline \hline
       & $S_{\nu}^{\rm tot}$ & $M^{\rm tot}$ & $N$  & $D_{\rm mean}$ & $S_{\nu}^{\rm mean}$ & $M^{\rm mean}$ \\
       &   Jy                          & \solm\            &          &  pc                    &  Jy                                & \solm\ \\
\hline
ALMA	& 0.52   &     53 &	12 &	0.025 &	0.042 &	4.42 \\
$\mu=2$ (x,y)	& 0.36    &    36	 &     12 &	0.013 &	0.026 &	2.76 \\
$\mu=2$ (x,z)	& 0.47    &    49	 &     12 &	0.017 &	0.039 &	4.1 \\
$\mu=2$ (y,z)	& 0.46     &   42	 &      8 &	0.018 &	0.050 &	5.2 \\
$\mu=200$ (x,y)	& 0.22 &       23 & 13	 & 0.015 & 	0.017 &	1.74 \\
$\mu=200$ (x,z)	& 0.24 &       25	 & 15	 & 0.014 & 	0.016 &	1.67 \\
$\mu=200$ (y,z)	& 0.28  &      24	 & 16 & 0.016 & 	0.021 &	2.19 \\
\hline
\end{tabular}
\end{center}
\end{table}

\end{document}